\documentclass{proceedingsM}
\usepackage{times,amsmath,amssymb}
\usepackage{graphicx,color}
\usepackage{hyperref}

\title{On a hierarchy of nonlinearly dispersive generalized KdV equations}
\author{Ivan C.\ Christov\footnotemark[1]%$ ^, $\footnotemark[2],
}
\address{{\footnotemark[1]} Theoretical Division and Center for Nonlinear Studies, Los Alamos National Laboratory, Los Alamos, NM 87545, USA}
\email{\href{mailto:christov@alum.mit.edu}{christov@alum.mit.edu}} %, email2@xyz.zyx.zz, email3@xyz.zyx.zz, }

\abstract{We propose a hierarchy of nonlinearly dispersive generalized Korteweg--de Vries (KdV) evolution equations based on a modification of the Lagrangian density whose induced action functional the KdV equation extremizes. It is shown that two recent nonlinear evolution equations describing wave propagation in certain generalized continua with an inherent material length scale are members of the proposed hierarchy. Like KdV, the equations from the proposed hierarchy possess Hamiltonian structure. Unlike KdV, however, the solutions to these equations can be compact (i.e., they vanish outside of some open interval) and, in addition, peaked. Implicit solutions for these peaked, compact traveling waves (``peakompactons'') are presented. %Using a well-chosen ansatz and the so-called variational approximation technique, we find explicit (albeit approximate) forml\ae\ for the traveling wave solution to equations from the proposed hierarchy.
}


\begin{document}

\maketitle

\section{Introduction}
\label{intro}

The Korteweg--de Vries (KdV) equation, originally derived to describe long waves over shallow water \cite{Boussinesq1877,Korteweg1895}, has emerged as a paradigm in the theory of solitons and integrable systems following the pioneering work of Zabusky and Kruskal \cite{Zabusky1965}. Beyond water waves, KdV arises in numerous other physical contexts \cite{Miura1976,Jeffrey1972}. Additionally, in modeling various physical phenomena beyond long waves over shallow water, many generalizations and modifications of the KdV equation have been derived over the years, including but not limited to: the modified KdV equation \cite{Miura1968,Miura1976}, the KdV--Burgers equation \cite{Jeffrey1972}, the cylindrical KdV equation \cite{Johnson1979}, the extended KdV equation \cite{Marchant1990}, the Camassa--Holm equation \cite{Camassa1993}, the $K(n,m)$ equations \cite{Rosenau1993}, the $K^*(l,p)$ equations \cite{Cooper1993}, the KdV--Kuramoto--Sivashinsky--Velarde equation \cite{Christov1995}, the generalized integrable KdV equation \cite[eq.~(4)]{Fokas1995}, KdV-based higher-order equations \cite{Salupere2001,Maugin2011}, hierarchical KdV equations \cite{Englebrecht2003,Randruut2009,Salupere2014}, the Destrade--Saccomandi equation \cite[eq.~(17)]{Destrade2006}, the Jordan--Saccomandi equation \cite[eq.~(4.14)]{Jordan2012} and $K(n,m)$ equations with non-convex advection \cite{Rosenau2014}. 

In the present work, we propose another hierarchy of generalized KdV equations based on modifying the Lagrangian density whose induced action functional is extremized by the KdV equation. Furthermore, we show that several members of this proposed hierarchy of nonlinearly dispersive evolution equations have been recently derived to describe wave propagation in generalized continua (of the type proposed by Rubin et al.~\cite{Rubin1995}) with an inherent material length scale. Finally, peaked, compact traveling wave solutions (``peakompactons'') of equations in the proposed hierarchy are constructed.

\section{A hierarchy of nonlinearly dispersive generalized KdV equations}

Let $\varphi = \varphi(x,t)$ be a field. In the present work, motivated by the form of the Lagrangian density whose corresponding action functional is extremized by the KdV equation (see, e.g., \cite[eq.~(2)]{Gardner1971}), we consider the class of Lagrangian  densities given by
\begin{equation}
	\mathcal{L}(\varphi;x,t) = \frac{1}{2}\varphi_x \varphi_t + \frac{1}{(n+2)(n+1)} (\varphi_x)^{n+2} - \frac{1}{m+1}(\varphi_{xx})^{m+1},\qquad n,m>0,
\label{eq:lagrangian}
\end{equation}
where subscripts denote differentiation with respect to an independent variable. For convenience, we have taken $n,m>0$ and, below, $n$ and $m$ typically take on integer values. However, the derivations below can be easily modified to hold for any real $n$ and $m$.
The corresponding Hamiltonian density can be obtained by the Legendre transformation (see, e.g., \cite[\S18]{Gelfand2000}):
\begin{equation}
\mathcal{H} = \frac{\partial \mathcal{L}}{\partial \varphi_t}\varphi_t - \mathcal{L} = -\frac{1}{(n+2)(n+1)} (\varphi_x)^{n+2} + \frac{1}{m+1}(\varphi_{xx})^{m+1}.
\label{eq:hamiltonian}
\end{equation}

The corresponding action functional $\int\mathrm{d}x\mathrm{d}t\,\mathcal{L}$ is extremized by requiring that its first variation vanish (see, e.g., \cite[\S11 and \S35]{Gelfand2000} for the details), namely
\begin{equation}
\frac{\partial \mathcal{L}}{\partial \varphi} - \frac{\partial}{\partial t}\left(\frac{\partial \mathcal{L}}{\partial \varphi_t}\right) - \frac{\partial}{\partial x}\left(\frac{\partial \mathcal{L}}{\partial \varphi_x}\right) + \frac{\partial^2}{\partial x^2}\left(\frac{\partial \mathcal{L}}{\partial \varphi_{xx}}\right) = 0.
\label{eq:gen-el}
\end{equation}
Substituting eq.~\eqref{eq:lagrangian} into eq.~\eqref{eq:gen-el}, assuming $\varphi$ is smooth enough so that $\varphi_{xt} = \varphi_{tx}$ and rearranging, we obtain the partial differential equation (PDE):
\begin{equation}
\varphi_{xt} + (\varphi_x)^{n}\varphi_{xx} + \left[(\varphi_{xx})^{m}\right]_{xx} = 0.
\label{eq:gkdv-ext-phi}
\end{equation}
Interpreting $\varphi$ as a potential, let $u(x,t) \equiv \varphi_x(x,t)$, then eq.~\eqref{eq:gkdv-ext-phi} becomes
\begin{equation}
u_t + u^{n}u_x + \left[(u_x)^{m}\right]_{xx} = 0,
\label{eq:gkdv-ext-u}
\end{equation}
giving the \emph{canonical form} of the proposed hierarchy of generalized KdV equations as parametrized by $(n,m)$.

\subsection{Conserved quantities}

One standard approach to constructing conserved quantities is to restrict to localized waveforms, specifically such that $\varphi \to \varphi_\pm$ ``sufficiently fast'' as $x\to\pm\infty$, where $\varphi_\pm$ are constants. In turn, this condition implies that $u \to 0$ ``sufficiently fast'' as $x\to\pm\infty$. Then, multiplying eq.~\eqref{eq:gkdv-ext-u} by $u^{k-1}$ for some integer $k$, integrating over $x\in(-\infty,+\infty)$ and performing some algebraic manipulations, leads to
\begin{equation}
\frac{\mathrm{d}I_{k}}{\mathrm{d}t}
%+\underbrace{\left.(k+1)\frac{u^{n+k+1}}{n+k+1}\right|_{-\infty}^{+\infty}}_{=0} 
+ k\int_{-\infty}^{+\infty}\mathrm{d}x\,\left[(u_x)^{m}\right]_{xx}u^{k-1} = 0,\qquad I_{k} \equiv \int_{-\infty}^{+\infty} \mathrm{d}x\, u^{k}.
\label{eq:int2}
\end{equation}
Note that for $k=1$, the integrand of the last term in eq.~\eqref{eq:int2}$_1$ is a complete differential, which when integrated vanishes by the specified asymptotic conditions, hence ${\mathrm{d}I_1}/{\mathrm{d}t} = 0$. For $k>1$, to establish conservation laws, the goal is to manipulate the integrand of the last term in eq.~\eqref{eq:int2}$_1$ into a complete differential. Performing integration by parts twice, it is readily shown that, for $k=2$, the term can be manipulated into a complete differential, hence ${\mathrm{d}I_2}/{\mathrm{d}t} = 0$. Meanwhile, for $k > 2$ (and $m\ne 0$) it can be shown that such a manipulation is impossible, hence no further conserved quantities of the form $I_k$ exist.

More generally, however, certain global conservation laws follow directly from the Hamiltonian structure of the proposed hierarchy of equations \cite{Maugin2002}. For example, the \emph{total wave energy}
\begin{equation}
H \equiv \int_{-\infty}^{+\infty}\mathrm{d}x\, \mathcal{H} \stackrel{\text{by}\;\eqref{eq:hamiltonian}}{=} \int_{-\infty}^{+\infty}\mathrm{d}x\, \left[ -\frac{1}{(n+2)(n+1)} u^{n+2} + \frac{1}{m+1}(u_{x})^{m+1}\right]
\end{equation}
is conserved, i.e., $\mathrm{d}H/\mathrm{d}t=0$.
Similarly, the \emph{total wave mass}
\begin{equation}
M \equiv \int_{-\infty}^{+\infty}\mathrm{d}x\, \frac{\partial \mathcal{L}}{\partial \varphi_t} \stackrel{\text{by}\;\eqref{eq:lagrangian}}{=} \frac{1}{2}\int_{-\infty}^{+\infty}\mathrm{d}x\, \varphi_x = \frac{1}{2}\int_{-\infty}^{+\infty}\mathrm{d}x\, u % = \frac{1}{2}[\varphi(+\infty) - \varphi(-\infty)].
\end{equation}
is conserved, i.e., $\mathrm{d}M/\mathrm{d}t=0$.
Likewise, the \emph{total wave momentum}
\begin{equation}
P \equiv -\int_{-\infty}^{+\infty} \mathrm{d}x\, \frac{\partial \mathcal{L}}{\partial \varphi_t}\varphi_x \stackrel{\text{by}\;\eqref{eq:lagrangian}}{=}  - \frac{1}{2} \int_{-\infty}^{+\infty} \mathrm{d}x\, (\varphi_x)^2 = -\frac{1}{2} \int_{-\infty}^{+\infty} \mathrm{d}x\, u^2
\end{equation}
is also conserved, i.e., $\mathrm{d}P/\mathrm{d}t=0$. Notice that $M = \tfrac{1}{2} I_1$ and $P = -\tfrac{1}{2} I_2$.

From Noether's theorem (see, e.g., \cite[\S4]{Kruskal1966}), these three physical conservation laws arise from underlying symmetries of the governing PDEs, specifically $P$, $H$ and $M$ are conserved due to translational invariance in space, translation invariance in time and the ability to add arbitrary constants to the potential $\varphi$, respectively, as can be easily  verified from eq.~\eqref{eq:gkdv-ext-phi}.

However, besides the special case $(n,m) = (1,1)$ (KdV), we do not expect members of this hierarchy to be ``fully integrable'' (i.e., to possess an infinite number of conservation laws), yet this remains an open question. An important step in this direction would be to establish a Miura transformation \cite{Miura1968} between the $(1,m)$ and $(2,m)$ equations in the proposed hierarchy for $m>1$.

\subsection{Scaling properties}

Above, we have implicitly assumed that we are working with dimensionless quantities, hence no coefficients involving typical physical parameters appear in the equations. Consider the scaling transformation
\begin{equation}
x\mapsto x/\ell,\qquad t\mapsto t/\tau,\qquad u \mapsto u/V,
\label{eq:rescale}
\end{equation}
where $\ell$, $\tau$ and $V$ are constants, which converts eq.~\eqref{eq:gkdv-ext-u} from its canonical form to a dimensional form:
\begin{equation}
u_t + \epsilon u^{n}u_x + \delta \left[(u_x)^{m}\right]_{xx} = 0,\qquad \epsilon \equiv \ell/(\tau V^{n}),\quad \delta \equiv \ell^{m+2}/(\tau V^{m-1}).
\label{eq:non-canonical}
\end{equation}
Clearly, given a non-canonical form of one of the equations of the hierarchy (with advective and dispersive coefficients $\epsilon$ and $\delta$, respectively), we can always choose characteristic scales $\ell$, $\tau$ and $V$ such that the rescaling  given in eq.~\eqref{eq:rescale} transforms the PDE \eqref{eq:non-canonical}$_1$ into its canonical given form in eq.~\eqref{eq:gkdv-ext-u}.

\section{Connection to other nonlinearly dispersive KdV-like equations from the literature}

Clearly, eq.~\eqref{eq:gkdv-ext-u} reduces to the KdV equation \cite[eq.~(1)]{Miura1968} for $(n,m)=(1,1)$, while it reduces to the \emph{modified} KdV equation \cite[eq.~(2)]{Miura1968} for $(n,m)=(2,1)$. Meanwhile, for $(n,m) = (1,3)$, eq.~\eqref{eq:gkdv-ext-u} takes the form of the unidirectional weakly-nonlinear equation for acoustic waves in an inviscid, non-thermally conducting compressible fluid with a material characteristic length coefficient that is a quadratic function of the velocity gradient, which was recently derived by Jordan and Saccomandi \cite{Jordan2012}. And, for $(n,m) = (2,3)$, eq.~\eqref{eq:gkdv-ext-u} takes the form of the unidirectional weakly-nonlinear equation for shear waves in a hyperelastic, incompressible solid with material dispersion (of the type suggested by Rubin et al.~\cite{Rubin1995}), which was derived by Destrade and Saccomandi \cite{Destrade2006}. The reader is referred to Maugin's historical review \cite{Maugin2011} for a discussion of various other KdV-like equations that arise in the theory of wave propagation in elastic solids.

Previously, Rosenau and Hyman \cite{Rosenau1993} proposed a different hierarchy of nonlinearly dispersive generalized KdV equations, namely the $K(n,m)$ equations:
\begin{equation}
u_t + (u^n)_x + (u^m)_{xxx} = 0.
\end{equation}
However, the $K(n,m)$ hierarchy is not Hamiltonian, thus lacking a physically desirable property that provides underlying geometric structure to many nonlinear evolution equations \cite{Holm2009}. A slight modification of the $K(n,m)$ hierarchy restores the Hamiltonian structure as shown by Cooper et al.~\cite{Cooper1993}, yielding the $K^*(l,p)$ hierarchy (with certain signs changes and setting $\alpha=1/2$ in \cite[eq.~(4)]{Cooper1993} for consistency of notation):
\begin{equation}
u_t + u^{l-2}u_x + u^p u_{xxx} + 2 pu^{p-1}u_x u_{xx} + \tfrac{1}{2}p(p-1)u^{p-2}(u_x)^3 = 0.
\label{eq:kslp}
\end{equation}
The corresponding Lagrangian density is $\mathcal{L}_{K^*}=\frac{1}{2}\varphi_x\varphi_t + \frac{1}{l(l-1)}(\varphi_x)^l - \frac{1}{2}(\varphi_x)^p(\varphi_{xx})^2$, which, when compared to eq.~\eqref{eq:lagrangian}, suggests that the present nonlinearly dispersive generalized KdV hierarchy is, in a sense, the ``most straightforward'' generalization of KdV's Hamiltonian structure. In keeping with the convention in the literature, we denote the hierarchy of equations  \eqref{eq:gkdv-ext-u} as $K^\#(n,m)$.

\section{Traveling wave solutions for the hierarchy}

Introducing the traveling wave ansatz $u(x,t) = U(\xi)$, where $\xi = x-ct$ for some wave speed $c>0$, reduces eq.~\eqref{eq:gkdv-ext-u} to an ordinary differential equation (ODE):
\begin{equation}
-cU' + U^{n}U' + \left[(U')^{m}\right]'' = 0,
\label{eq:tws-ode}
\end{equation}
where primes denote differentiation with respect to $\xi$.
A first integral of eq.~\eqref{eq:tws-ode} is evident by inspection,
%\begin{equation}
%-cU + \frac{1}{n+1}U^{n+1} + \left[(U')^{m}\right]' = C_1,
%\label{eq:tws-ode-1}
%\end{equation}
%where $C$ is a constant of integration. 
then rearranging the latter, multiplying by $U'$ and rewriting all terms as complete derivatives immediately yields a second integral:
\begin{equation}
(U')^{m+1} = C_2 + C_1 U + \frac{(m+1)}{2m}c U^2 - \frac{(m+1)}{(n+1)(n+2)m}U^{n+2},
\label{eq:tws-ode-2}
\end{equation}
where $C_1$ and $C_2$ are constants of integration. Finally, eq.~\eqref{eq:tws-ode-2} can be solved implicitly:
\begin{equation}
\int \frac{\mathrm{d}U}{\left[C_2 + C_1 U + U^2\left(\kappa - \gamma U^{n}\right)\right]^{1/(m+1)}} = \pm\xi + \xi_0,\qquad \kappa\equiv \frac{(m+1)}{2m}c,\quad \gamma\equiv \frac{(m+1)}{(n+1)(n+2)m},
\label{eq:tws-ode-3}
\end{equation}
where we choose the positive root on the left-hand side of eq.~\eqref{eq:tws-ode-3}$_1$, hence  the $\pm$ sign on the right-hand side.

A complete study of eq.~\eqref{eq:tws-ode-3} for various boundary conditions is beyond the scope of the present work.  
%As mentioned above, the (trivial) special case(s) $(n,m)\in\{-2,-1\}\times\{-1,0\}$ are to be treated separately. 
Nevertheless, for $(n,m)=(1,1)$ (KdV), solutions can be written \emph{explicitly} using hyperbolic or elliptic special functions \cite{Miura1976,Jeffrey1972}. Some solutions are also known for $m=3$, and an intriguing aspect of the latter is that they are \emph{compact}, i.e., $u$ vanishes outside of some open $x$-interval. Following \cite{Destrade2006,Jordan2012}, let us restrict to localized waveforms such that $u,u'\to0$ as $|x|\to\pm\infty$, then $C_1=C_2=0$. It follows that the denominator of the integrand on the left-hand side of eq.~\eqref{eq:tws-ode-3}$_1$ vanishes at $U=\left\{0,U_\text{max}\right\}$, where $U_\text{max}\equiv [(n+1)(n+2)c/2]^{1/n}$. The Lipschitz condition is not satisfied at these points and different solutions can be pieced together there, allowing for \emph{compactification} of the waveform. Notice that $U=0$ is always a solution of eq.~\eqref{eq:tws-ode-2} for $C_2=0$.

Now, recognizing the integral on the left-hand side of eq.~\eqref{eq:tws-ode-3}$_1$ (with $C_1=C_2=0$) as a hypergeometric function \cite[\S16.15]{DLMF}, we obtain
\begin{equation}
\left(\frac{m+1}{m-1}\right) U(\kappa U^2)^{-1/(m+1)}
   \,{}_2F_1 \left[\frac{1}{m+1},\frac{m-1}{(m+1)n};1+\frac{m-1}{(m+1)n};\frac{\gamma}{\kappa}U^n\right] = \pm\xi + \xi_0.
\label{eq:tws-ode-4}
\end{equation}
where the last integration constant $\xi_0$ is fixed by requiring that $U(0)=U_\text{max}$:
\begin{equation}
\xi_0 = \left(\frac{m+1}{m-1}\right) U_\text{max} \left(\kappa U_\text{max}^2\right)^{-1/(m+1)} 
   \, {}_2F_1\left[\frac{1}{m+1},\frac{m-1}{(m+1) n};1+\frac{m-1}{(m+1) n};1\right].
\label{eq:xi0}
\end{equation}
As in \cite{Destrade2006,Jordan2012}, it is now straightforward to piece together a compact traveling wave solution such that $U=0$ for $\xi \in (-\infty,-\xi_0]\cup[+\xi_0,+\infty)$, $U$ is given implicitly by eqs.~\eqref{eq:tws-ode-4}--\eqref{eq:xi0} for $\xi \in (-\xi_0,+\xi_0)\backslash \{0\}$ and $U(0) = U_\text{max}$. 

It can be verified that $U$ and $U'$ are continuous at $\xi=\{0, \pm\xi_0\}$, while
\begin{equation}
U'' = U^{-\frac{m-3}{m+1}}  \left(U_\text{max}^n - U^n\right)^{-\frac{m-1}{m+1}} \left(c(n+1)-U^n\right) (U_\text{max}^n)^{-\frac{2}{m+1}} \kappa^{-\frac{m-1}{m+1}}(n+2) c/(2m),
\end{equation}
which suggest there are several possibilities for $U''$. The term $U^{-\frac{m-3}{m+1}}$ dominates the behavior of $U''$ as $U\to0^+$ ($\xi\to\pm\xi_0$) with $U''\to 0$ for $m<3$, $U''\to \sqrt{c/6}$  for $m=3$ (see also \cite[eq.~(4.18)]{Jordan2012}) and $|U''|\to\infty$ for $m>3$. Similarly, the term  $(U_\text{max}^n - U^n)^{-\frac{m-1}{m+1}}$ dominates the behavior of $U''$ as $U\to U_\text{max}^-$ ($\xi\to0$) and, since the compact solutions derived hold only for $m>1$ and $n>0$, $|U''|\to\infty$ as $U\to U_\text{max}^-$ ($\xi\to0$).

The presence of an infinite second derivative at $\xi = 0$ means that the compact waveforms constructed are \emph{peaked} like the (non-compact) \emph{peakon} solutions of the Camassa--Holm equation \cite{Camassa1993}, hence we refer to the family of solutions given by eqs.~\eqref{eq:tws-ode-4}--\eqref{eq:xi0} by the portmanteau \emph{peakompactons}. The possibility of an infinite second derivative as $\xi\to\{0,\pm\xi_0\}$ implies that we have constructed \emph{pseudo-classical} solutions of eq.~\eqref{eq:gkdv-ext-u}, i.e., piecewise solutions that satisfy eq.~\eqref{eq:gkdv-ext-u} in a classical sense away from $\xi=\{0,\pm\xi_0\}$, and, moreover, lead to well defined limits for the nonlinearly dispersive terms $m(m-1)(U')^{m-1}(U'')^2$ and $m(U')^{m-1}U'''$ as $\xi\to\{0,\pm\xi_0\}$ (see, e.g., the illuminating discussion in \cite{LiOlver1997,LiOlver1998}). 

\begin{figure}[!h]
\centerline{\includegraphics[width=0.75\textwidth]{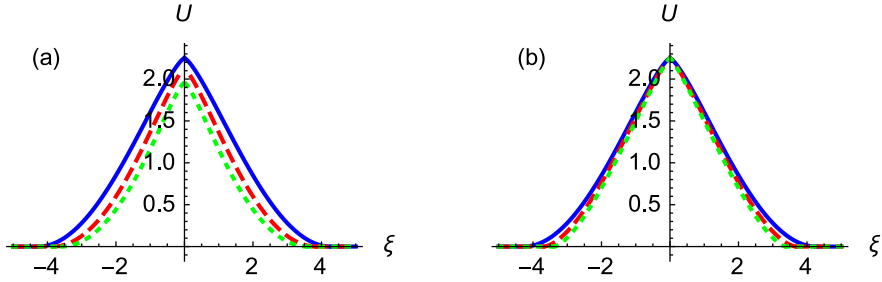}}
\caption{Illustration of the $K^\#(n,m)$ hierarchy's peakompacton solutions given by eqs.~\eqref{eq:tws-ode-4}--\eqref{eq:xi0} for $c=0.75$ and (a) $n=1,2,3$ (solid, dashed, dotted) and $m=3$, (b) $n=1$ and $m=3,4,5$ (solid, dashed, dotted).}
\label{fig:pk}
\end{figure}

The peakompacton solution \eqref{eq:tws-ode-4}--\eqref{eq:xi0} is illustrated in fig.~\ref{fig:pk} for various values of $(n,m)$. Note that $U_\text{max}$ is independent of $m$, hence the form of the nonlinear dispersion has no effect on the peak height. However, both $n$ and $m$ affect the length of the finite support since $\xi_0$ is a function of both. Similarly, $c$ is featured in both the expression for $U_\text{max}$ and $\xi_0$, thus the wave speed affects both the width and height of the peakompactons.

\section{Conclusion}

In the present work, we proposed a hierarchy of nonlinearly dispersive generalized Korteweg--de Vries (KdV) evolution equations based on a modification of the Lagrangian density whose induced action functional is extremized by the KdV equation. Conservation laws, scaling properties and peaked, compact traveling wave solutions for this class of PDEs were discussed. It is important to emphasize that the proposed $K^\#(n,m)$ hierarchy of nonlinearly dispersive KdV-like PDEs possesses compact traveling wave solutions, like the $K(n,m)$ hierarchy. However, unlike the $K(n,m)$ hierarchy, the $K^\#(n,m)$ equations also possess Hamiltonian structure. Indeed, the $K^*(l,p)$ hierarchy restored the Hamiltonian structure to the $K(n,m)$ equations, however, to the best of our knowledge, none of the equations from the $K^*(l,p)$ hierarchy have been derived from physical principles, i.e., these equations remain phenomenological. On the other hand, at least two of the equations from the proposed $K^\#(n,m)$ hierarchy (with $m>1$) have been  independently derived to describe wave propagation in generalized continua.

In the present work, our goal was to present the $K^\#(n,m)$ class of nonlinear evolution equations and some of their elementary properties. The mathematical analysis of these PDEs remains to be done. A number of questions were noted in the text above: e.g., Miura transformations between members of the class, integrability of the hierarchy and/or individual members of it, existence of other (exotic) exact solutions, collision properties of the known solitary waves, etc. However, of specific interest would be to understand whether eq.~\eqref{eq:gkdv-ext-u} with $m$ even (e.g., 2) has a physical interpretation, since (so far) only eq.~\eqref{eq:gkdv-ext-u} with $m$ odd (specifically, $1$ and $3$) has been derived in physical contexts.

\section*{Acknowledgments}
I.C.C.\ was supported by the LANL/LDRD Program through a Feynman Distinguished
Fellowship. Los Alamos National Laboratory (LANL) is operated by Los Alamos National Security, L.L.C.\ for the National Nuclear Security Administration of the US Department of Energy under Contract No.\ DE-AC52-06NA25396. 
Prof.~Andrus Salupere and the local organizing and international scientific committees of the 2014 IUTAM Symposium on Complexity of Nonlinear Waves in Tallinn, Estonia are acknowledged for their work in making the meeting a great success. The symposium's hospitable intellectual environment made the present work possible.
Finally, I.C.C.\ would also like to thank F.\ Cooper, J.M.\ Hyman, P.M.\ Jordan, T.\ Kress, A.\ Oron and A.\ Saxena for helpful discussions on the topic of the present work.

% BibTeX users please use
%\bibliographystyle{unsrt}
%\bibliography{file}   % name your BibTeX data base

\end{document}